\documentclass[%
 reprint,
superscriptaddress,
 amsmath,amssymb,
 aps,pre,hyphens,floatfix
]{revtex4-2}

\usepackage[]{amsmath}
\usepackage{mathrsfs}
\usepackage{mathtools} 
\usepackage{enumerate}
\usepackage[usenames,dvipsnames,svgnames,table]{xcolor}
\usepackage{IEEEtrantools}
\usepackage{graphicx}
\usepackage{array}
\usepackage{multirow}
\usepackage{verbatim}
\usepackage{soul}
\usepackage{tabularx}
\usepackage{booktabs,siunitx}
\usepackage[T1]{fontenc}
\usepackage{hyperref}
\usepackage{siunitx}
\usepackage{wrapfig}
\usepackage{algorithm}
\usepackage{algpseudocode}
\usepackage{indentfirst}
\usepackage{makecell}

\newcommand{\mycomment}[1]{}

\newcommand{\av}[1]{\langle {#1} \rangle}
\newcommand{\stk}[1]{\ifmmode\text{\sout{\ensuremath{#1}}}\else\sout{#1}\fi}

\begin{document}

\preprint{APS/123-QED}

\title{Improving power-grid systems via topological changes \\ or how can self-organized criticality help power-grids}


\author{G\'{e}za \'{O}dor}
\email[]{odor@mfa.kfki.hu}
\affiliation{Institute of Technical Physics and Materials Science, Centre for Energy Research, P.O. Box 49, H-1525 Budapest, Hungary}


\author{Istv\'an Papp}
\affiliation{Institute of Technical Physics and Materials Science, Centre for Energy Research, P.O. Box 49, H-1525 Budapest, Hungary}

\author{Krist\'of Benedek}
\affiliation{Budapest University of Technology and Economics, Műegyetem rkp. 3, H-1111 Budapest, Hungary}

\author{B\'{a}lint Hartmann}
\email[]{hartmann.balint@ek.hun-ren.hu}
\affiliation{Institute of Energy Security and Environmental Safety, Centre for Energy Research, P.O. Box 49,
 H-1525 Budapest, Hungary}
 
\date{\today}

\begin{abstract}
Cascade failures in power grids occur when the failure of one component or subsystem causes a chain reaction of failures in other components or subsystems, ultimately leading to a widespread blackout or outage. Controlling cascade failures on power grids is important for many reasons like economic impact, national security, public safety and even rippled effects like troubling transportation systems. Monitoring the networks on node level has been suggested by many, either controlling all nodes of a network or by subsets.  
This study identifies sensitive graph elements of the weighted European power-grids (from 2016, 2022) by two different methods. Bridges are determined between communities and "weak" nodes are selected by the lowest local synchronization of the swing equation. In the latter case we add bypasses of the same number as the bridges at weak nodes, and we compare the synchronization, cascade failure behavior by the dynamical improvement with the purely topological changes. The results are also compared if bridges are removed from networks, which results in a case similar to islanding, and with the addition of links at randomly selected places. Bypassing was found to improve synchronization the best, while the average cascade sizes are the lowest with bridge additions.
However, for very large or small global couplings these network changes do not help, they seem to be useful near the synchronization transition region, where self-organization drives the power-grid. Thus, we provide a demonstration for the Braess' Paradox on continent-sized power grid simulations and uncover the limitations of this phenomenon. 
We also determine the cascade size distributions and justify the power-law tails near the transition point on these grids.

\end{abstract}

\maketitle

\section{Introduction\label{sec:1}}

Blackouts and other failures frequently occur in stressed electrical power 
systems with low operational margins. Therefore, they have to adapt to changes in the use of electrical energy. 
Earlier power-grids were not originally designed for deregulated markets 
that appeared in the 1990s, these grids transmitted large amounts of electrical 
power across interconnections. 
As both the grid and the operators were unable to handle fast-developing disturbances, the result was a global increase in major blackouts. 
Nowadays, power industry has been addressing decarbonization needs with a large integration of renewable generation and electrification, which introduces strong fluctuations.
Furthermore, power systems are significantly affected by the increasing number of climate change
induced extreme weather~conditions~\cite{Nauck_2023}. Thus, it is necessary to redesign the transmission and distribution grids to address these changes.

Modelling blackouts and other failures is a great challenge, which has been attempted by various approximations. Earlier ones used a direct current (DC) approach, similar to sand-piles \cite{Car,Car2,Bis} or fiber boundle like \cite{dou_robustness_2010,Yagan} models. These provided heavy tailed, power-law (PL) statistics of blackout sizes similar to observations \cite{Car, dobson2007c} via self-organized criticality (SOC)~\cite{bak1987}.That is generated by the competition of supply and demand, tuning power systems to a critical point, where the PL-s occur.

However, SOC is not the only possible mechanism suggested to describe PL-s. The highly optimized tolerance (HOT) model is also proposed~\cite{PhysRevE.60.1412}, but it is
probably more appropriate to describe certain types of failures without cascades~\cite{outagecikk}. The spectral analysis of outage duration times suggests that there exist cases for which HOT is more applicable. It has been
proposed~\cite{PhysRevE.60.1412} that the competition of service capacity and failures can also lead to SOC in a reaction-diffusion type model, leading to PL distributed repair times. City size PL distribution, corresponding to power load, has also been hypothesed to be a possible reason for the PL outage cascades \cite{nesti2020}.

Later, alternating current (AC) models appeared, by solving the swing equation~\cite{fila} equivalent to the second order Kuramoto equation \cite{kura}, set up for the phases of the voltages with the addition of some threshold criterion of line failures~\cite{SWTL18,M2,M3,M4,M5,USAEUPowcikk}. 
As it is difficult to solve these nonlinear equations for large systems, linearization has also been used frequently~\cite{kaiser2021net}.
Another major challenge is the heterogeneity of power-grids systems, which has been considered in various ways, see discussion in~\cite{heterogeneity}. 

Predicting, controlling and avoiding blackouts \cite{10164641,M2} and helping to design more error prone networks and methods has been the subject of many other studies, see for example~\cite{Scholl2016}.
Most of them use the above approximations and try to isolate most vulnerable points, by a frequency  analysis of solitary nodes~\cite{Olmi-Sch-19}.

Here we contribute to the modelling of cascades based on the complete Kuramoto equations, without any linearization, by numerical simulations on large European high voltage (HV) power-grids. We investigate effects of changing network topology on the synchronization and cascade failure dynamics in comparison of several ways. The first path is purely static, based on the network community analysis and aims to determine the effects of addition or removal of bridges between communities.
The second path is dynamic, it uses the solution of the swing equations and identifies the nodes of lowest local synchrony. Bypassing these nodes, via the addition of edges of the same number as the bridge links, we can compare the methods. It seems that generally it is possible to get some gain over the purely static topological extensions. We also compare these results with the random addition of links. 

Power-grid extensions require large investments and are supposed to make the system operation more robust. Yet, counter-intuitively, increasing the capacity of existing lines or adding new lines may also reduce the overall system performance and even promote blackouts due to Braess’ paradox~\cite{Braess1968, braess2005paradox}. 
Braess’ paradox was theoretically modeled~ \cite{Cohen1991, Witthaut2012, Witthaut2013, Fazlyaba2015, nagurney2016observation, Coletta2016, motter2018antagonistic}, but has not yet been proven in realistically scaled power grids.
Very recently a topological theory has been provided that reveals the key mechanism and predicts Braessian grid extensions from the network structure and a linearized power flow DC approximation~\cite{Sch_fer_2022}. Now, we extend this study by our dynamic AC analysis, suggesting that the Braess's paradox does not show up near the synchronization transition, where self organization drives the power-grid system. Recently, a similar conclusion has been drawn concerning the usefulness of islanding~\cite{USAEUPowcikk}, i.e. improved stability following failure cascades in the neighborhood of the synchronization transition and deficiency, away from the transition region. 

It was also argued in recent years that (N-1) congestions are a direct consequence of the topological and reactance structure of the power grid and how these interact with the loadability of the lines. As highlighted in ~\cite{9855886}, the imbalances in the reactance structure of the grid are a leading cause of congestions, and using the technique of shadow capacity analysis, strengthening of the grid can be carried out in a way to avoid poor powerflow redistributions after an outage. These findings are also underpinned by the results of the present paper.

\section{Methods and models \label{sec:2}}

\subsection{Solving the massive Kuramoto synchronization equations}

The time evolution of power-grid synchronization is described by the swing 
equations~\cite{swing}, set up for mechanical elements (e.g.~rotors in generators 
and motors) with inertia.
It is formally equivalent to the second-order Kuramoto equation~\cite{fila}, 
for a network of $N$ oscillators with phases $\theta_i(t)$.
Here we use a more specific form~\cite{Olmi-Sch-19,POWcikk,USAEUPowcikk,heterogeneity},
which includes dimensionless electrical parametrization and approximations for unknown ones:
\begin{equation}\label{kur2eq}
{\ddot{{\theta }}}_{i}+\alpha {\ }{\dot{{\theta}}}_{i}=P_i
+\frac{{P}_{i}^{max}}{{I}_{i}{\ }{\omega }_{G}}{\ }\sum
_{j=1}^{N}{{W}_{\mathit{ij}}{\ }\sin \left({\theta }_{j}-{\theta
}_{i}\right)} \ .
\end{equation}
In this equation $\alpha$ is the damping parameter, which describes 
the power dissipation, or an instantaneous feedback~\cite{Powfailcikk}, 
we keep $K:=P_i^{max}$ as a global control parameter, related to the 
maximum transmitted power between nodes, $I_i=I$ inertia and $\omega_G$ system frequency  are kept constants in the lack of our knowledge; and $W_{ij}$, is the adjacency matrix of the network, which contains admittance elements, calculated from impedances as described in~\cite{heterogeneity}.
The quenched external drive, denoted by $P_i:=\omega_i^0$, which is 
proportional to the self-frequency of the $i$-th oscillator, 
carries a dimension of inverse squared time $[1/s^2]$, and describes 
the power in/out of a given node, when Eq.~(\ref{kur2eq}) corresponds to the swing equation (phases without amplitudes) of an AC power circuit.

Here, as commonly done with the first-order Kuramoto model, the self-frequencies are drawn from a zero-centered Gaussian random 
variable, as the rescaling invariance of the equation allows to transform it within a rotating frame. 
For simplicity, one can assume that $\omega_i(0)$ is drawn from 
the same distribution as $\omega_i^0$ and numerically set
$\omega_i(0)=\omega_i^0$, amounting to taking $[s]$=1.
In our present study, the following parameter settings were used:
the dissipation factor $\alpha$ is chosen to be equal to $0.4$ to
meet expectations for power grids, with the $[1/s]$ inverse time 
physical dimension assumption, but we also tested the $\alpha=3.0$
case, which can describe a system with stabilizing linear feedback
\cite{POWcikk,USAEUPowcikk}.

To solve the differential equations, in general we used the adaptive Bulirsch-Stoer stepper~\cite{boostOdeInt}, which provides more precise
results for large $K$ coupling values than the fourth-order Runge--Kutta method.The nonlienearity introduces chaotic 'noise', 
even without stochasticity, thus a de-synchronization transition occurs by lowering $K$. The solutions also depend on the actual
quenched $\omega_i^0$ self-frequency realization. To obtain reasonable fluctuations of the averages of 
measured quantities, we needed strong computing resources, using parallel codes running on GPU HPC machines. 
To obtain stronger synchronization solutions, the initial state was set to be phase synchronized: $\theta_i(0)=0$, but 
due to the hysteresis, one can also investigate other uniform random distributions like: $\theta_i(0) \in (0,2\pi)$.
The initial frequencies were set to be: $\dot{\theta_i}(0)=\omega_i^0$.

To characterize the phase transition properties, both the phase
order parameter $R(t)$ and the frequency spread $\Omega(t)$, 
called the frequency order parameter, were studied. 
We measured the Kuramoto phase order parameter:
\begin{equation}\label{ordp}
z(t_k) = r(t_k) \exp\left[i \theta(t_k)\right] = 1 / N \sum_j \exp\left[i \theta_j(t_k)\right] \ .
\end{equation}
Sample averages for the phases
\begin{equation}\label{KOP}
R(t_k) = \langle r(t_k)\rangle
\end{equation}
and for the variance of the frequencies
\begin{equation}\label{FOP}
        \Omega(t_k,N) = \frac{1}{N} \langle \sum_{j=1}^N (\overline\omega(t_k)-\omega_jt_k))^2 \rangle
\end{equation}
were determined, where $\overline\omega(t_k)$ denotes the mean frequency within each respective sample at time step
$t_k = 1 + 1.08^{k}$, $k=1,2,3...$.
Sample averages were calculated for solutions with hundreds of independent self-frequency realizations for each control parameter, while for determinig PDF-s of the failure cascades about 20.000 samples were used to estimate the histograms.

\subsection{Cascade failure simulations}

We have extended the numerical solution of the Kuramoto equations with a threshold dynamics, such that in case of an overflow of power on the edges, we removed them during the simulation of a cascade failure. This method is similar to the one published in \cite{SWTL18,USAEUPowcikk}. 
Following a thermalization, which is started from a phase ordered state and line-cuts are not allowed, we perturbed the system by removing a randomly selected link, in order to simulate a power failure event. Following that, if the ensuing power flow on a line between neighboring nodes was greater than a threshold:
\begin{equation}\label{Fij}
        F_{ij} = | \sin(\theta_j-\theta_i) | > T\,,
\end{equation}
so that that line is regarded as overloaded,
we removed this link from the graph permanently and measured the total number of line failures $N_f$ of the simulated blackout cascades of each realizations, corresponding to different
$\omega_i(0)$ self frequency values.
Finally, we applied histogramming to determine the PDFs of $N_f$.
In the vicinity of criticality, one usually expects power-law distributions of the form
\begin{equation}
        p(N_f)\sim N_f^{-\tau}\, ,
        \label{Nfpdf}
\end{equation}
thus we plotted our results on the log-log scale.

\subsection{The power-grid networks}

In this study,  various modifications of European power-grids introduced in a previous work~\cite{heterogeneity} were investigated. These are the EU16 (European 2016), EU22 (European 2022) graphs, for which the backbone of the used network data is from the \href{https://www.power.scigrid.de/pages/downloads.html}{SciGRID project}, which relies on the statistics of ENTSO-E and data obtained from OpenStreetMap (.osm) files. These contain information on the topology, geographical coordinates of nodes, and lengths, types, voltage levels of cables. Since acquiring data from .osm files is not always possible, the resulting data set may be incomplete. To resolve the problem, we made assumptions in~\cite{heterogeneity} to substitute the missing data in order to obtain fully weighted networks. We used the giant component of the networks, giving $N=$ \num{13420} nodes linked with $L=$ \num{17749} edges for EU16 grid and $N=$ \num{7411} nodes connected by $L=$ \num{10298} for the EU22 network.  We have also performed graph theoretical analysis on them to determine graph invariants and their community structure. The resulting graphs are summarized in the Table \ref{tab:comms}.
\begin{table}[ht]
\centering
\begin{tabular}{ccccc}
\toprule
Community &  \makecell{Size \\ (EU22)} & \makecell{$\langle k \rangle$ \\ (EU22)} & \makecell{Size \\ (EU16)} & \makecell{$\langle k \rangle$ \\ (EU16)}  \\
\midrule
         1 &   924 & 2.72 & 4285 & 2.83  \\
         2 &   479 & 2.70 & 2526 & 2.66  \\
         3 &  2016 & 2.84 & 1527 & 2.67 \\
         4 &   698 & 3.06 & 1461 & 2.72 \\
         5 &   595 & 2.94 & 1455 & 2.69  \\
         6 &  1059 & 2.66 & 966 & 2.77 \\
         7 &  1237 & 2.68 & 638 & 2.57 \\
         8 &    16 & 2.81 & 289 & 2.06 \\
         9 &   332 & 2.18 & 277 & 2.99 \\
        10 &    55 & 2.74 & 26 & 3.07 \\
        11 & - & - &  22 & 3.31 \\
        12 & - & - &  6 & 2.66 \\
\bottomrule
\end{tabular}
\caption{Community sizes and average degrees for different data-sets, for the resolution $\Gamma=10^{-4}$.
We refer to sizes here as number of nodes in the respecting community and provide their average degree.}\label{tab:comms}
\end{table}
As we can see, the EU22 network has a lower number of communities, nodes and links. Other graph measures, like the degree and cable lengths distributions also suggest that the EU22 is incomplete, but still provides an excellent possibility to study the effects of the network topology on the synchronization dynamics~\cite{heterogeneity}

\subsection{Creation of bridges between communities\label{sec:3}}

Detecting communities in networks aims to identify groups of nodes in the network that are more densely connected to each other than to the rest of the network.
While several clustering methods exist, they split into hierarchical and non-hierarchical methods. Hierarchical methods build a hierarchy of communities by recursively dividing 
the network into smaller and smaller subgroups, while non-hierarchical methods directly assign nodes to communities.

For detecting the community structure, we chose the hierarchical Louvain \cite{Blondel2008} method for its speed and scalability. This algorithm runs almost in linear time on sparse graphs, therefore, it can be useful on generated test networks with increased size. It is based on modularity optimization. 
The modularity quotient of a network is defined by \cite{Newman2006-bw}
\begin{equation}
Q=\frac{1}{N\av{k}}\sum\limits_{ij}\left(A_{ij}-\Gamma
\frac{k_i k_j}{N\av{k}}\right)\delta(g_i,g_j),
\end{equation}
the maximum of this value characterizes how modular a network is. Here  $A_{ij}$ is the weighted adjacency matrix, containing the admittances calculated in~\cite{heterogeneity}. 
Furthermore, $k_i$, $k_j$ are the weighted node degrees of $i$ and $j$ and $\delta(g_i,g_j)$ is $1$, when nodes $i$ and $j$ were found to be in the same community, or $0$ otherwise. $\Gamma$ is the resolution parameter, which allows a more generalised community detection, merging together smaller communities.

In network analysis, a bridge (Br) refers to a link or an edge that connects nodes from different communities or components of a network. Bridges are crucial, because they establish connections between otherwise separate parts of a network, facilitating the flow of information or influence between different communities. Removing bridges can lead to a fragmentation of the network into isolated components. In the EU16 network, we selected 1250 bridges of all communities from the "true" communities detected, where we optimized the modularity at  $\Gamma = 1$, with Leiden \cite{TraagWaltmaVanEck2018_LouvainLeiden} algorithm we did not find better results, separation was worse with 449 communities, connected by 1281 bridges. In the case of the EU22 network, for $\Gamma = 1$ we found 94 communities, connected by 507 bridges.

To increase stability, we applied simple duplication of bridges. Alternatively, we also tried to remove almost all bridges between the communities, which leads to an "islanded" graph, where cascade failures are more localized.
To avoid working with a fully disconnected network, we removed bridges randomly, starting with the biggest number of communities. We continued removing bridges until the network still remained connected (Br-).
\begin{figure}
    \centering
    \includegraphics[width=\columnwidth ]{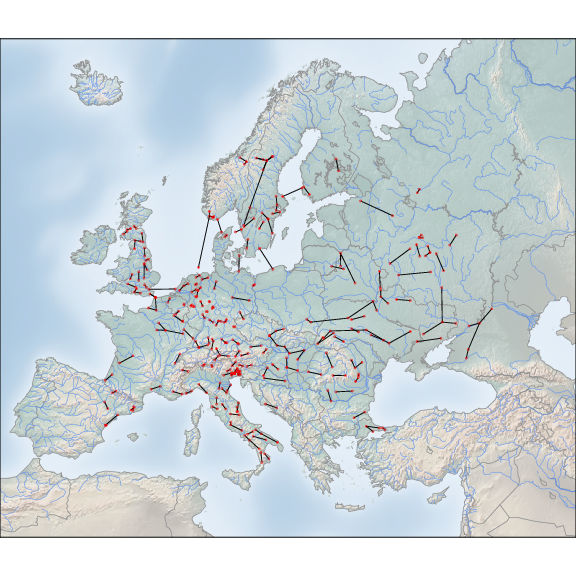}
    \caption{Here we show the (red) nodes that have been selected with the bypass cutoff method and happen to be on the (black) bridge edges between communities. These represent 10$\%$ of selected nodes by this method.}
    \label{fig:12comm}
\end{figure}
Results of these network decompostions are published in~\cite{heterogeneity}.

\subsection{Creation of bypasses at weak nodes of the local Kuramoto solution\label{sec:4}}

We performed dynamic stability analysis of the network, by identifying weak nodes via the local order parameter, defined as
\begin{equation}\label{eq:loc_ord_param}
	        r_i(t)= \frac{1}{N_{\mathrm{i.neigh}}}
	        \left|\sum_j^{N_{\mathrm{i.neigh}}}
		A_{ij} e^{i \theta_j(t)}\right|. 
\end{equation}
This method is a bit more precise than just finding the solitary nodes of outstanding frequencies~\cite{Olmi-Sch-19}, as it is based on a large ensemble average and considers 
interactions with the neighboring nodes.  

Having the weak nodes of the grid identified, we propose a way to improve the stability by creating some extra links, called \textit{bypass}es (Bp), which interconnect the critical points of the network, hence making the graph more robust. While there are several ways to achieve this, we present one of the most simple ones: by creating so-called triangles or doubling links between weak neighbors. Both methods are used in actual power grid development to increase the redundancy of supply.

In graph-theoretical language, a \textit{triangle} is composed of three nodes, each connected by links to the other two.
In a mathematical sense, triangles are used to calculate the \textit{global clustering coefficient}, which characterizes the robustness of the network:
\begin{equation}
    C = \frac{3\times\mathrm{number\; of\; triangles}}{\mathrm{number\; of\; all\; triplets}},
\end{equation}
The motivation behind creating triangles is to increase the robustness of the network by enhancing the clustering coefficient.

Using Eq.~(\ref{eq:loc_ord_param}), we group the nodes into different synchronization categories, and by selecting the worst synchronized group, we can implement a so-called "bypass algorithm". The algorithm does the following:
it goes through the list of the worst synchronized nodes. For each weak node, it checks the neighborhood. If in the neighborhood there is a weak node, it doubles the link between them. The doubled link inherits parameters of the original one. If there is no other weak node in the neighborhood, it chooses the two closest neighbor nodes and connects them with a new link, creating a triangle. The new link inherits the average parameter values of the links connecting to the selected weak node.

\begin{figure}[h]
    \centering
    \includegraphics[width=0.6\columnwidth]{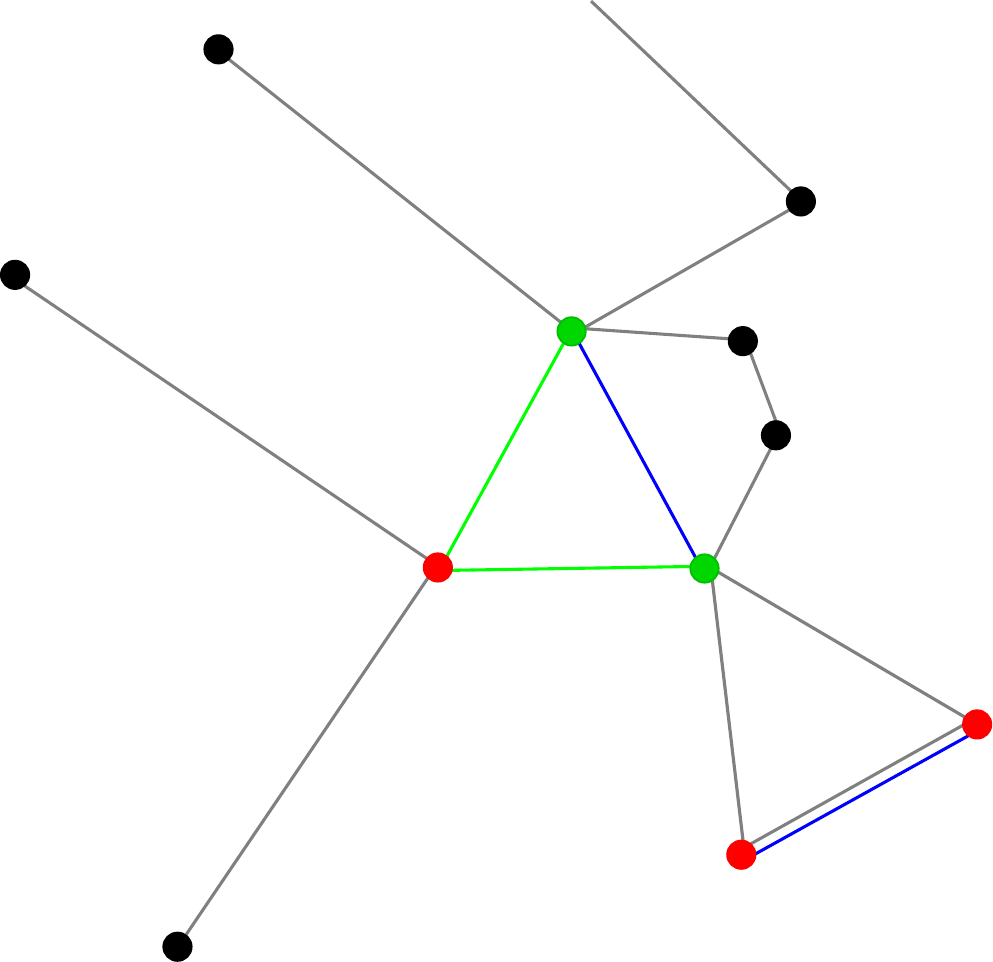}
    \caption{Sketch of the bypass algorithm on a small sample. We colored with red the badly synchronized nodes, obtained by the local Kuramoto order parameter $r$, with green the two closest neighbors, connecting to the weakest node. Going through the red (weakest) nodes we perform the following "bypass algorithm": either create a triangle with the help of the two closest nodes, which are not weak (green triangle, with one blue edge), if in the neighborhood there is no other weak node. If there are two neighboring weak nodes, we double their connecting edge, (grey link doubled with a blue). The blue links mark the newly added edges to the network.}
    \label{fig:bpalg}
\end{figure}

These newly added links are expected to increase the resilience of the network, since if one of the routes gets cut, there will be alternative paths for the energy transfer through the weak nodes, avoiding overloads and line failures. On the other hand, it is possible to have regions of parameters when this interconnectedness turns into a disadvantage and causes larger cascade failures. Based on simulation results, this typically happens at lower $\mathrm{K}$ values (see Fig. \ref{fig:Cut-ave}) or larger $\alpha$ values in the still relatively small $\mathrm{K}$ region (see Fig. \ref{fig:Cut-ave3})

\begin{figure}[h]
    \centering
    \includegraphics[width=\columnwidth ]{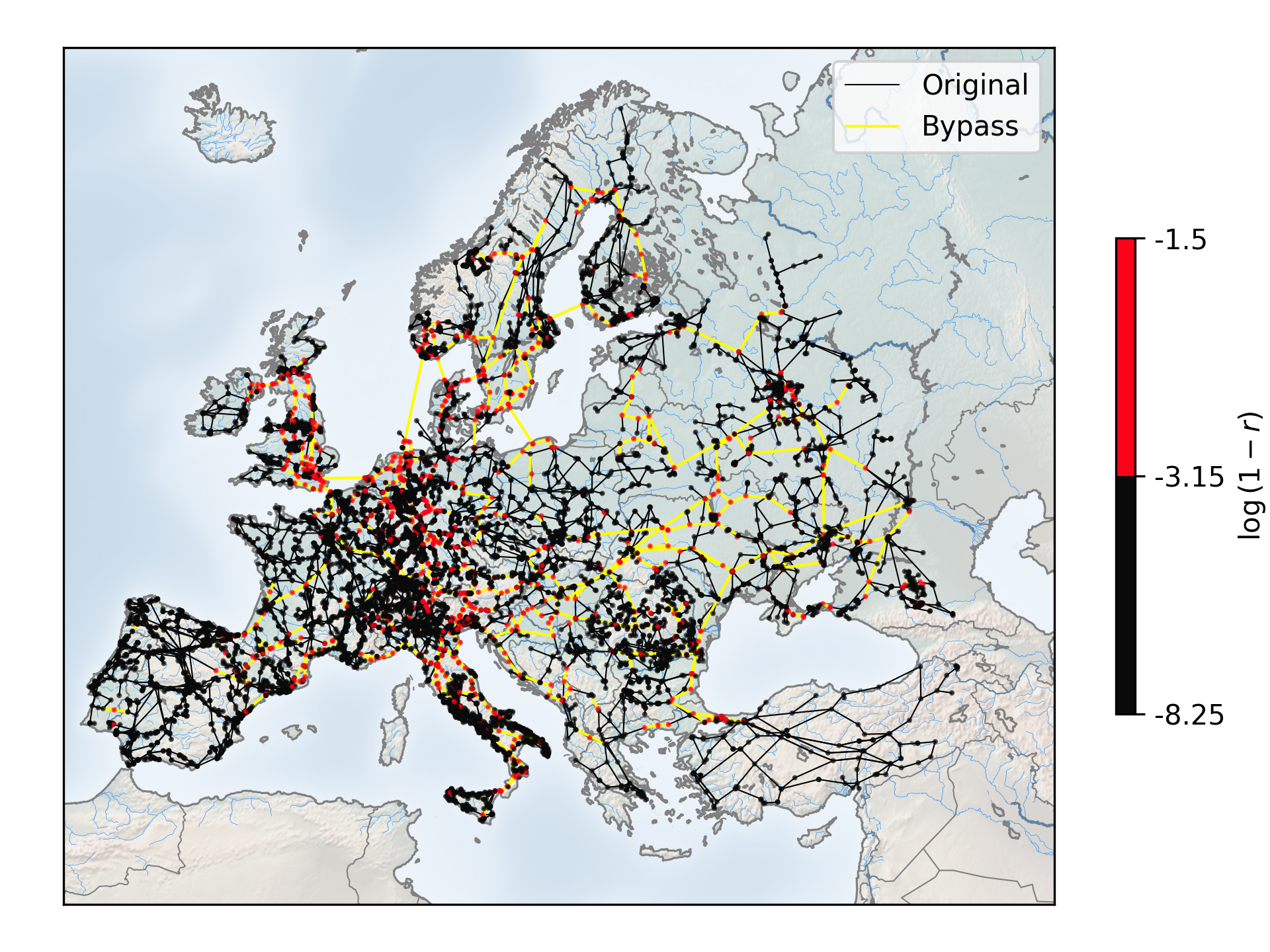}
    \caption{New links are added to the EU-HV 2016 network, denoted by yellow lines, using the bypass algorithm, where low local synchronization is obtained by the solving Eq.(\ref{kur2eq}).  $r_i$ is encoded by the colors. Red dots are the weakest nodes.  }
    \label{fig:eu16_bp}
\end{figure}

The results for the bypassing are shown in Fig.~\ref{fig:eu16_bp}, with the yellow links, denoting the newly added components to the original network edges, colored by black. Choosing the bins of the local synchronization, i.e. the $\log{(1-r)}$ values properly, we can set by a threshold how many new links we add to the network. This makes the method more general and comparable with the bridge and community analysis, where the topology of the network is given, and we cannot control the newly added components, except by modifying the modularity resolution $\Gamma$ of the community detection.

\section{Results} 

\subsection{Comparison of topological changes}

\begin{figure}
    \centering
    \includegraphics[width=\columnwidth ]{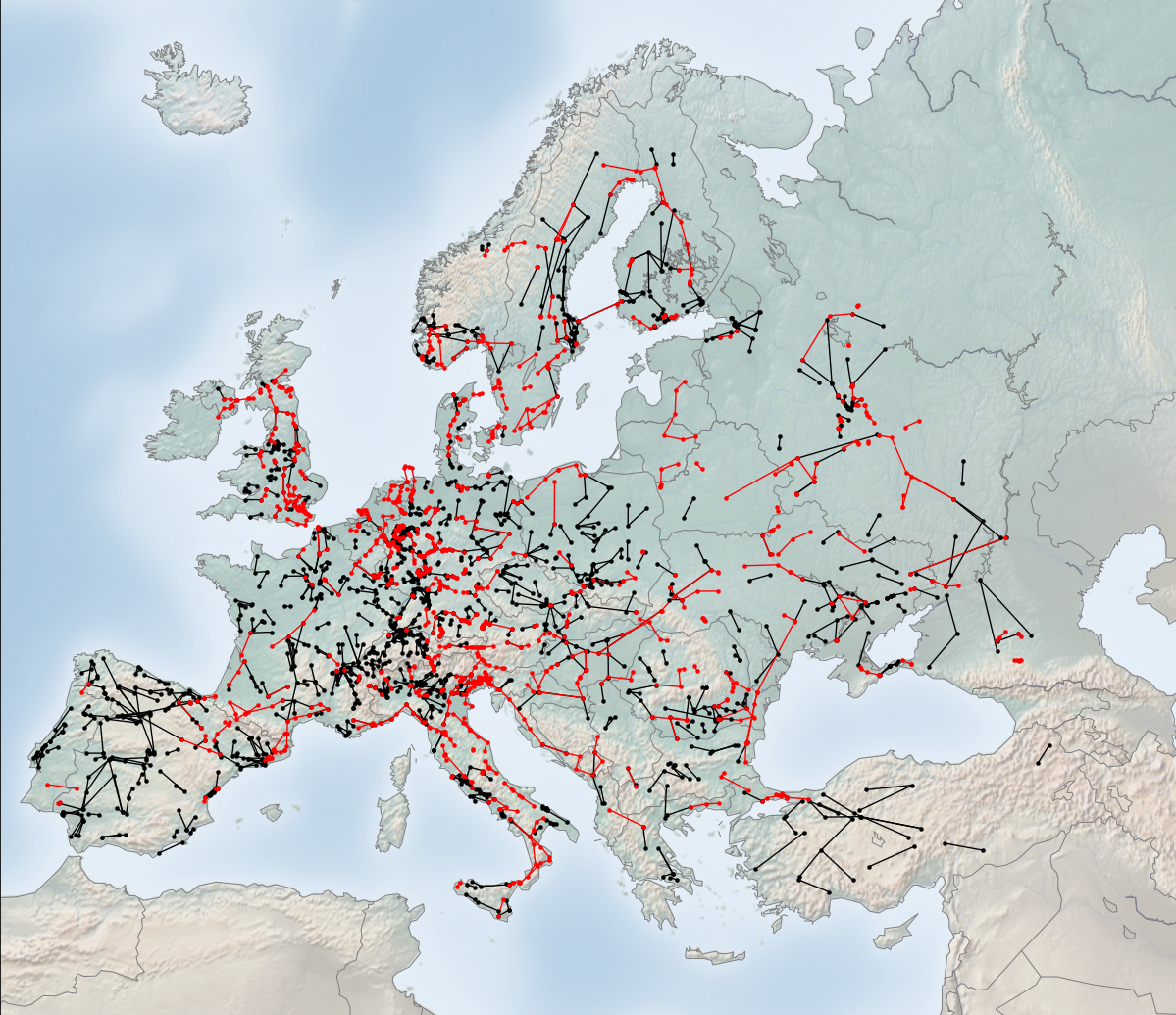}
    \caption{Here we show the difference between the set of (red) nodes, links selected with the bypass method for EU16 network and 
    the set of black nodes, links that are on the bridges. The edges were inserted between the nodes selected by the bypass method, increasing the network's modularity score.}
    \label{fig:12comm-diff}
\end{figure}

To see the differences between the new links, which were added by the static bridges and the dynamically determined bypasses, we have plotted their
overlaps in Fig.~\ref{fig:12comm} and differences in Fig.\ref{fig:12comm-diff}.
Red links correspond to links where bypasses are added but no bridges, while the black ones to the opposite
We can see that the red links are concentrated in the middle of Europe, dividing East and West, and in the UK. The black ones are mainly on the Iberian Peninsula, France and Ukraine among smaller communities, obtained by $\Gamma=1$.

\subsection{Results for phase, frequency and cascade sizes}

\subsubsection{EU16 results}

\begin{figure}
    \centering
    \includegraphics[width=\columnwidth ]{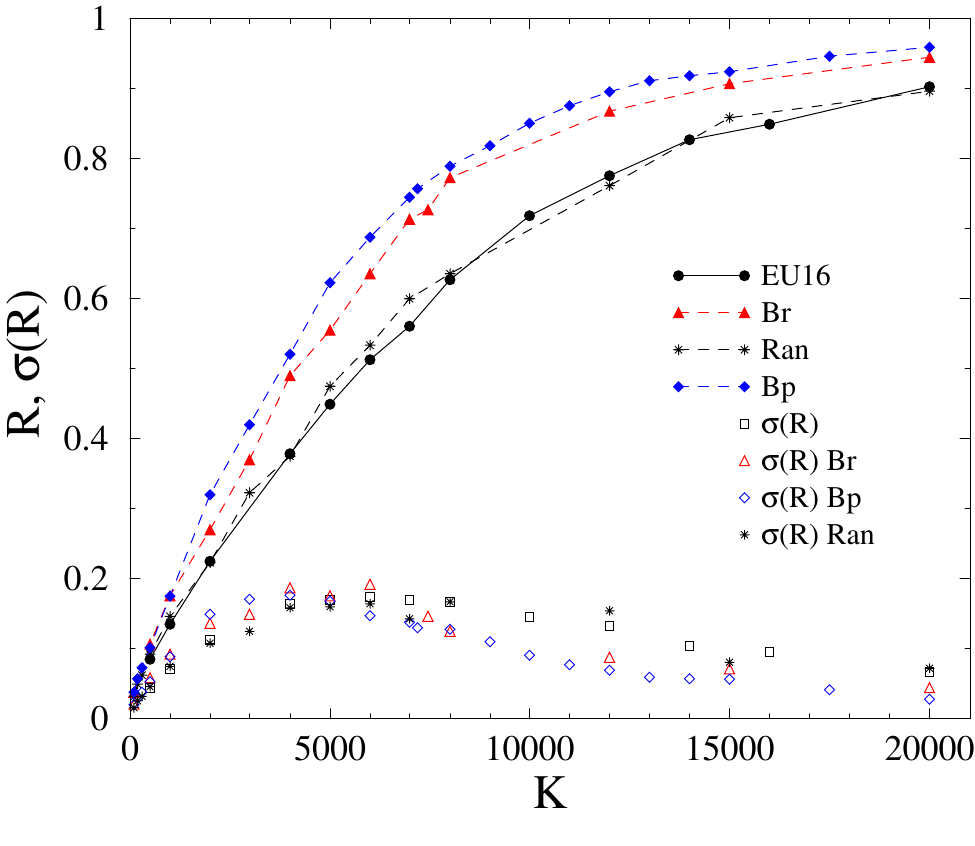}
    \caption{Comparison of numerical solutions of $R$ and its variance $\sigma(R)$ at the
     end of the thermalization in the steady state, for the original (EU16), 
     randomly extended (Ran), bridged (Br) and bypassed (Bp) EU16 networks at $\alpha=0.4$. }
    \label{fig:R-K}
\end{figure}

We have calculated the synchronization stability measures following a thermalization process with $t_{Th}=2000$ and after that, by allowing cascade failures for $t_{Cut}=1000$ iterations. We started the systems from phase synchronized states, for different global coupling $K$ values by solving the swing equations (\ref{kur2eq}), to achieve higher synchronization as in the case of the second order Kuramoto equation, an initial condition
dependent, hysteretic behavior occurs. Thus if the simulation starts from a state with random phases, the solver arrives at lower
$R$ and higher $\Omega$ steady state values, and the transition point shifts to higher $K$-s.
We utilized the adaptive Bulrich-Stoer stepper, because the synchronization transition happens at large $K$ values. 
Averaging has been done for $100 - 1000$ initial random Gaussian self-frequency distributions as well as via temporal averaging in the last decades of the steady states. Fig.\ref{fig:R-K} shows that the Kuramoto order parameter $R$ increases slowly from zero to $\simeq{1}$ for the weighted, randomly supplemented, bridged, bypassed and truncated cases. The highest synchronization values are obtained around $K_c\simeq 6000$, in the original, weighted case, where the fluctuations have a peak, marking the neighborhood of a SOC state.
The gain in $R$ is the best $\simeq 60\%$ for the bypassed case, which was obtained by strengthening the weakest points in the graph, as discussed before. But the bridged network also shows a considerable increase in $R$ near $K_c$: $\simeq 50\%$.

We have not found such an improvement for the global frequency spreading 
order parameter~\ref{FOP} $\Omega$, as shown on Fig. ~\ref{fig:O-K} in the Supplementary Material. The differences are small among the modified networks and the original one for the whole $K$ parameter space scanned. But this may not mean that local improvements are not possible, i.e. the zero centered Gaussian initial $\omega(i,0)$-s can split to multi centered distributions by slightly shifted peaks, as the empirical data of \cite{Oberhofer_2023,Olmi-Sch-19} show. Maybe, this kind of topological supplementation is not as effective as islanding of certain weak domains. 

We have also compared the average cascade sizes using $T=0.99$ and found $\simeq 50\%$ smaller blackouts near $K_c\simeq 6000$ between
the original and bridged cases, as shown in Fig. ~\ref{fig:Cut-ave} in the Supplementary Materials.
Here we used $t_{max}=1000$ time steps for the maximum size of observation of cascades following the initialization and initial random line cuts. The bridged case provides the smallest cascade sizes, but the bypassed case also improves the results with respect
to the original network in the region $20 < K < 20.000$.
However, far from the synchronization transition region, i.e. for
$K<20$ or $K>20.000$, there is no such benefit. In fact, the original
networks perform better for small global couplings (total transmitted power).
This is the Braess's Paradox, which is even more visible for $\alpha=3$ in Fig.~\ref{fig:Cut-ave3} in the Supplementary Material. But in real power-grids, due to an SOC mechanism, systems operate near the 
synchronization transition, where this phenomenon does not seem to occur.

Probability distributions of the $N_f$ are also shown in the inset of Fig. \ref{fig:Cut-ave} slightly above $K_c$, at $K=7000$ and $K=7500$,
where occurrence of heavy tails can be observed, similarly to an unweighted network~\cite{USAEUPowcikk}. We fitted the tails by PL-s
for $N>100$, resulting in a decay exponent: $\tau_t\simeq 2.6$, somewhat
larger than for unweighted networks~\cite{Powfailcikk,USAEUPowcikk}. 
\begin{figure}
    \centering
    \includegraphics[width=\columnwidth ]{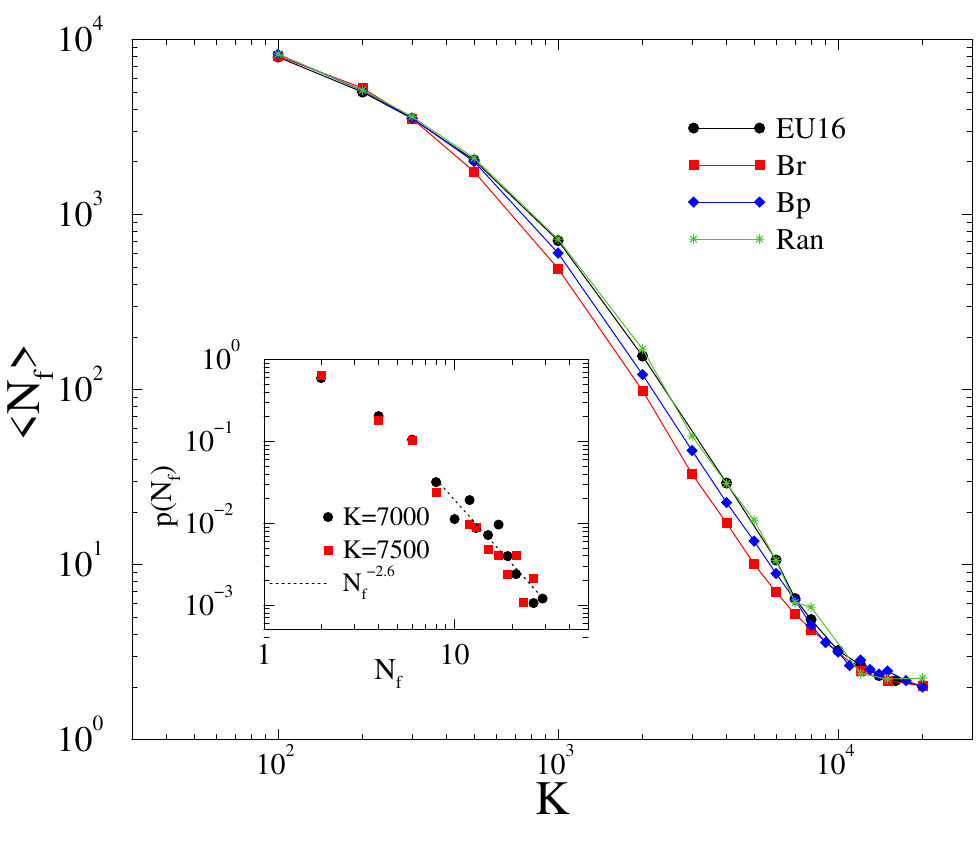}
    \caption{Comparison of dynamic simulation results of the average cascade
     sizes $\langle N_f\rangle$ at $T=0.99$ for the original, randomly extended, 
     bridged and bypassed EU16 networks, using $\alpha=0.4$. 
     Note the crossings of lines for small and large $K$-s, corresponding to Braess's paradox. Inset: PDF of line-cuts for $T=0.99$, $K=7000$ (circles) and $K=7500$ (boxes), 
     dashed line: PL fit for $N_f>10$.}
    \label{fig:Cut-ave}
\end{figure}

These findings have been investigated further for $\alpha=3$ as in previous publications~\cite{Powfailcikk,USAEUPowcikk}, corresponding to larger dissipation, or equivalently to instantaneous feedback mechanisms.
Fig. \ref{fig:Cut-ave3} in the Supplementary Material depicts effects of network extensions are much more pronounced than by $\alpha=0.4$. 
Again, the bypassed case provides the best performance for phase synchronization stability, after thermalization or after the end of the cascade. The cascade size distribution exhibits fat tails at $K\simeq K_c=12000$ and at the threshold $T=0.99$, which can be fitted by a PL for $N>3$ with an exponent $\tau_t \simeq 2.6$, similarly to $\alpha=0.4$.
In the original network we can see some improvement, similar to what is seen on the unweighted EU16 network~\cite{USAEUPowcikk}, which we attributed to islanding effects.

We have also simulated an almost complete islanding, by removing most of the 1250 bridges without cutting the network's full integrity. AsThe remaining, islanded network shows very low levels of $R$ in the steady state. 

As the Kuramoto order parameter changes, the addition of bridges move the $\sigma(R)$ peaks towards lower couplings, meaning the global synchronization occurs at lower couplings.

We can also see improvements in the frequency spread results in Fig.\ref{fig:O-K3}, as shown in the Supplementary Material, except for the bridge removal. The best global $\Omega$-s can be achieved by the
bridge additions, followed by the bypass technique.
We do not find improvements following the cascades, the diluted networks exhibit larger $\Omega$-s.

But the average cascade sizes can be improved a lot by the addition of bypasses or bridges \ref{fig:Cut-ave3}.
The bridges seem to be the most efficient for $\langle N\rangle$, followed by the bypasses
in the synchronization transition region: $30 < K < 20.000$. Below $K=30$ we can see crossing lines, corresponding the change of the tendencies as for $\alpha=0.4$. Rather large cascades appear for the truncated network, so this kind of islanding truncation method does not increase network stability.

\subsubsection{EU22 results}

\begin{figure}
    \centering
    \includegraphics[width=\columnwidth ]{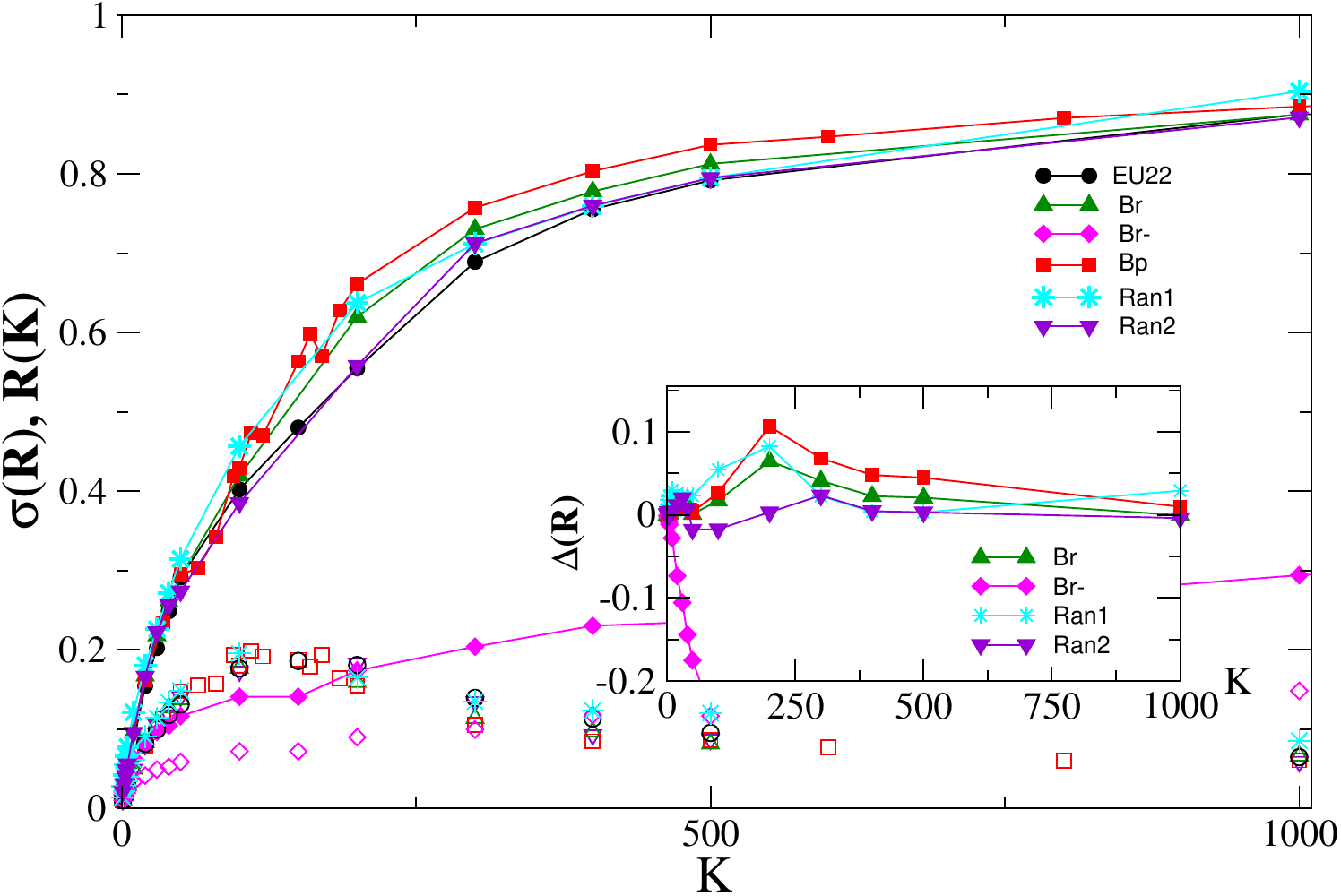}
    \caption{Comparison of numerical results of $R$ and $\sigma(R)$ at the end of the thermalization in the steady state for the original (EU222), two randomly extended versions (Ran1, Ran2), bridged (Br) and bypassed (Bp) EU22 networks, using $\alpha=3.0$. The inset shows the $R$ deviations on different networks from the original. Improvement is the best near the synchronization transition $K_c\simeq 150$, especially for bypasses.}
    \label{fig:22-R-K-a3}
\end{figure}

To test the robustness of the results, we repeated the analysis done in the previous section for the EU22 network. Here we show the results for $a=3.0$ only, for the $a=0.4$ case they are similar, but with smaller deviations between different networks solutions 
As the EU22 network is smaller than the EU16, the differences between the $R(K)$ results are smaller. Thus, we also show $\Delta(R) = R(K)_{mod.} - R(K)_{org.}$ in the inset of Fig.~\ref{fig:22-R-K-a3}. Again the bypassed network is the most stable, with about a 10\% increase near the transition point $K_c\simeq 125$. The synchronization transition points can be read-off by the peaks of $\sigma(R)$ on the graph, they do not seem to depend too much on the network version.
The bypassed network result is followed by the bridged case, with a 5\% maximum increase in $R(K_c)_{mod.}$, while the random edge addition hardly provides any improvement. 
By removing bridges between communities, except a few to maintain single connectedness,
the remaining network becomes very unstable, as indicated by the magenta diamond symbols.

The advance of the bypassed network over the others becomes more visible in the case of $\Omega$ at $\alpha=3.0$ (see Fig.~\ref{fig:22-O-K-a3} in the Supplementary Material). The frequency deviations remain small, and the bridge removal increases the spread.

\begin{figure}[H]
    \centering
    \includegraphics[width=\columnwidth ]{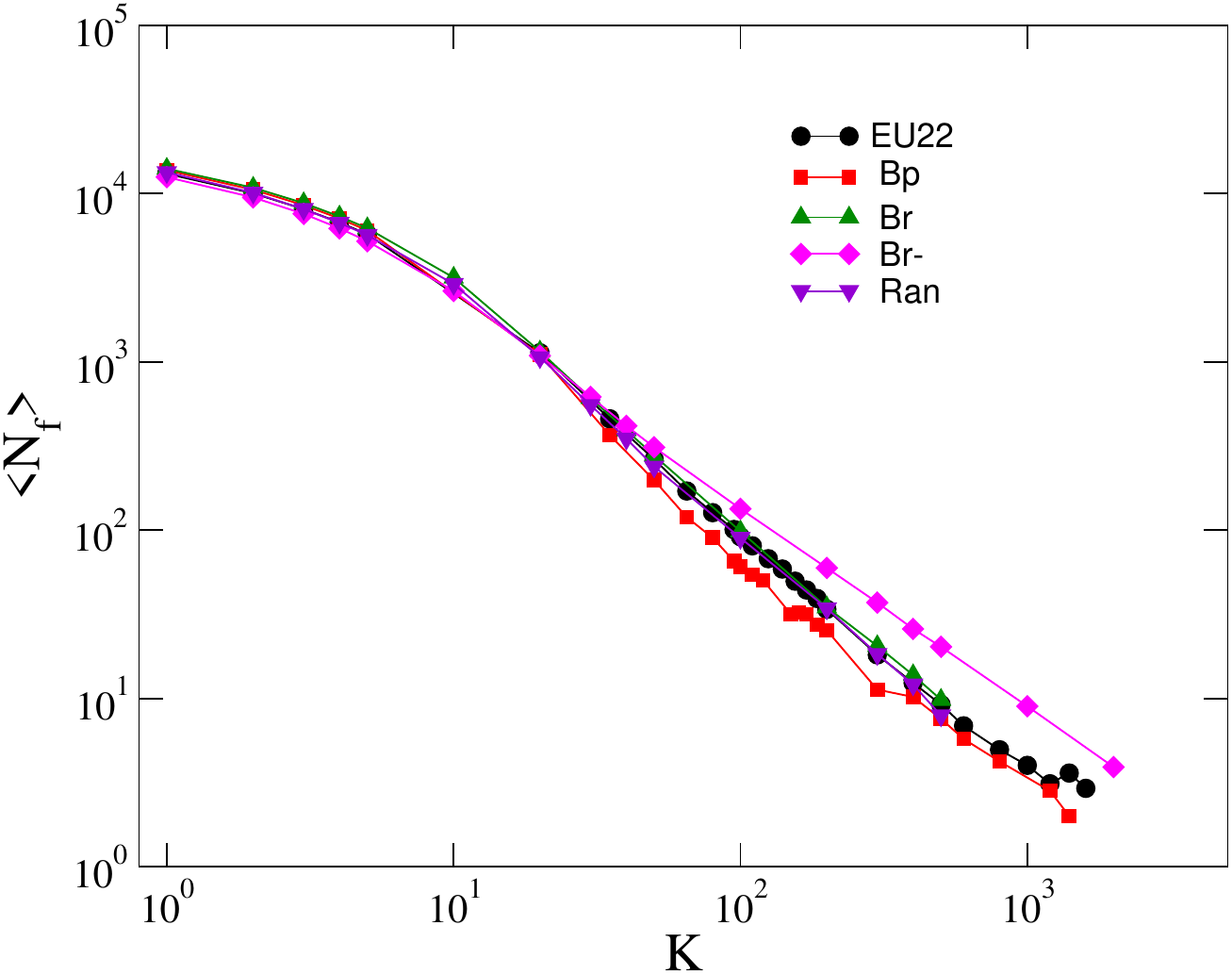}
    \caption{Comparison of dynamic simulation results of the average cascade sizes $\langle N_f\rangle$, for $T=0.99$ for the original, randomly extended, bridged and bypassed EU22 networks, using $\alpha=3.0$. Bypassed results are the best, while bridge removal results are the least favourable.}
    \label{fig:22-Cut-ave3}
\end{figure}

The average cascade sizes also show similar trends for the EU16 grids, but now the bypassed setting
proves to be the best. The worst scenario arises if bridges are removed (Fig.\ref{fig:22-Cut-ave3}).

\section{Conclusions}

This study extended former large scale EU power-grid and cascade simulations~\cite{USAEUPowcikk} 
by edge weights, as described in ~\cite{heterogeneity}.
Different network topology optimization strategies were compared for the EU16 and EU22 HV power-grids. The addition of the same number of new edges provided the best improvements of the global phase synchronization by dynamically obtained bypasses at the weak nodes. The efficiency of this enhancement is followed by the static bridge additions. For the cascade sizes, the bridge method proved to be the winner for the EU16 network.

These improvements are effective in the middle range couplings, near the synchronization point, where these systems presumably tune themselves by self-organization. This is similar to our findings for the usefulness of islanding~\cite{USAEUPowcikk}.
What can be the reason behind this phenomenon? Clearly phase chaoticity is maximal near the synchronization transition, which can help redistributing the power from local overloads and increase the stability against cascade failures.
Investigating further the effects of such 'noise' would be the target of further research. Our present results also provide a possible range of control parameters, where the Braess's paradox may take place. Further network modification methods, like the introduction of DC lines\cite{DC-control} or consideration of different voltage amplitudes could also be interesting research directions.

We have confirmed again, as in~\cite{USAEUPowcikk,Powfailcikk}, that near the synchronization point, PL distributed cascade sizes occur, in agreement with the historical data. We have also verified the usefulness of the network improvement results with the random link additions and the bridge removals, which provided clearly much worse stability and cascade size behaviors.

This work provides possibilities of further generalization and may help designers for improving the next generation of power-grids. One of the most striking conclusions is that systems benefit from SOC not only
by optimizing resource allocation, but an increased stability against failures and network changes appears
in the optimal control parameter ranges.

Another valuable finding is that while the addition of links is generally considered to strengthen the structure of the grid, such additions not only have to consider the topological parameters, but the electric characteristics of power lines as well. While improper placement of new lines may help to optimize power flows through the grid, in the case of outages they could also be the driving force towards cascading failures. The understanding of these mechanisms requires the use of heterogeneous and weighted network representations, which will stay in the focus of our future work as well.

\begin{acknowledgments}
Support from the Hungarian National Research, Development and Innovation Office NKFIH (K128989) and from the ELKH grant SA-44/2021 is acknowledged. We thank KIFU for the access to the national supercomputer network, Jeffrey Kelling for developing the GPU HPC code and Shengfeng Deng for his comments and the maintenance of our local computing resources. 
\end{acknowledgments}

\bibliography{topology}

\section{Supplementary material}

\begin{figure}
    \centering
    \includegraphics[width=\columnwidth ]{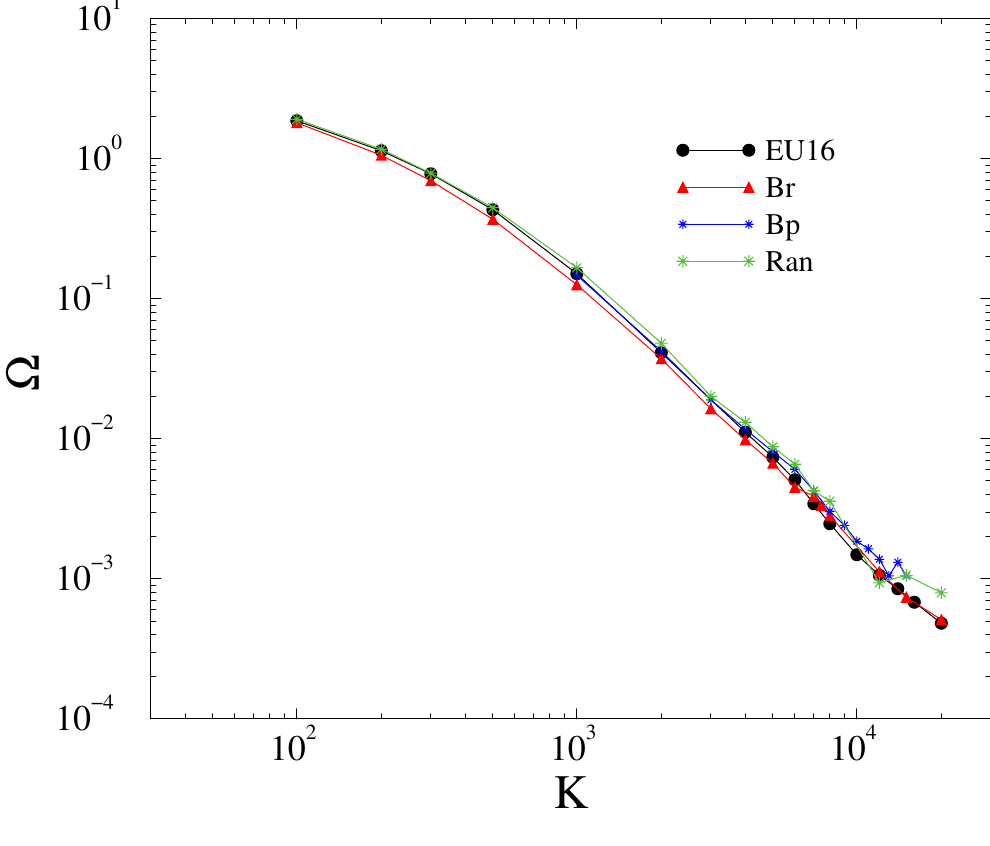}
    \caption{Comparison of numerical results for $\Omega$ at the end of the thermalization in the steady state for the original (EU16), randomly extended (Ran), bridged (Br) and bypassed (Bp) EU16 networks, 
    using $\alpha=0.4$.}
    \label{fig:O-K}
\end{figure}

Here we provide more figures depicting the results. Fig.\ref{fig:O-K} illustrates the dependency of frequency spread on $K$ in EU15 networks. The differences between the various scenarios are negligible.
As one see the differences are negligible.

\begin{figure}
    \centering
\includegraphics[width=\columnwidth ]{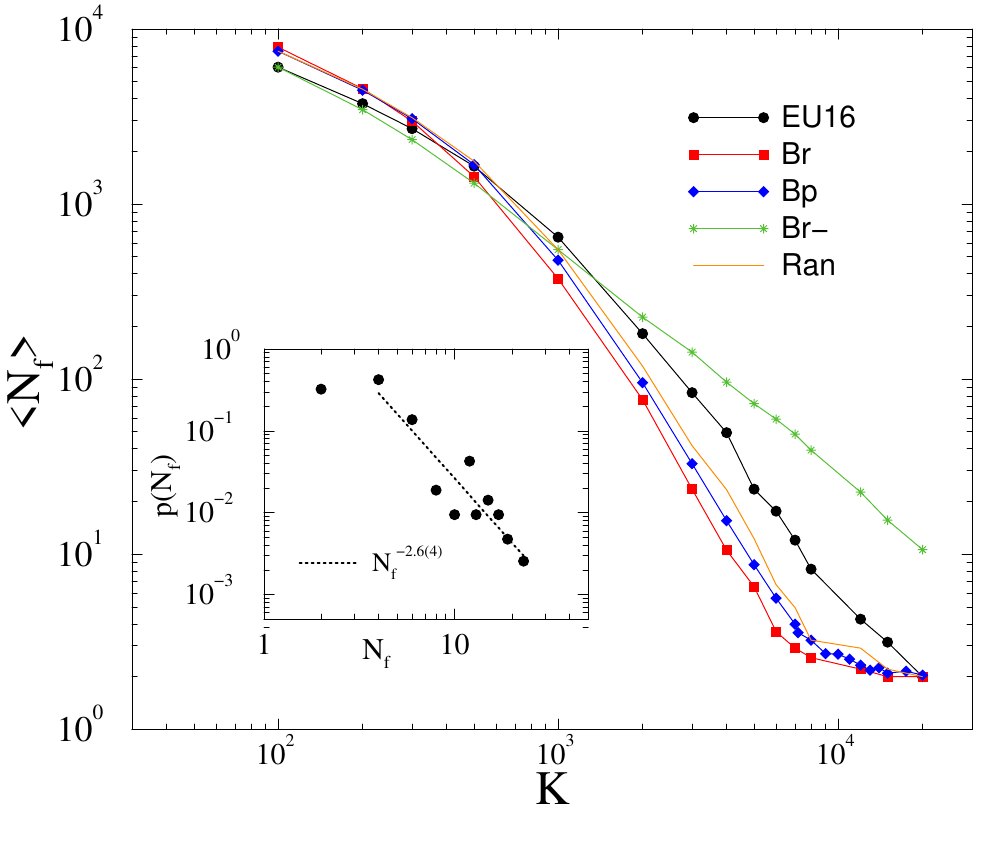}
    \caption{Comparison of dynamic simulation results of the average cascade sizes $\langle N\rangle$, for $T=0.99$ for the original, randomly extended (Ran), 
     bridged (Br) and bypassed (Bp) networks EU16, using $\alpha=3.0$. The Braess's Paradox appears for very small $K<40$ and very large global couplings.
     Inset: PDF of line-cuts for $T=0.99$, $K=12000$ (circles), 
     dashed line: PL fit for $N > 3$.}
    \label{fig:Cut-ave3}
\end{figure}

The next plot Fig. \ref{fig:Cut-ave3} shows the average cascade sizes for $\alpha=3$, as well as the PDF of cascades near the synchronization point: $K=12000$ and $T=0.99$. The bridged network provides the smallest cascades for $500 < K < 20000$, with a $\simeq 90\%$ gain as the best. This is followed by the bypassed network by a $\simeq 80\%$ gain, while the random addition of edges also provides a $\simeq 70\%$ improvement. The results $\langle N\rangle$ for the bridge removed, almost islanded case become much larger than the original, if $K > 500$. For $100 \le K \le 500$, the enforced networks exhibit larger cascades, demonstrating the Braess' effect, but a slight improvement occurs in the bridge truncated case,  suggesting that islanding effects emerge.

\begin{figure}
    \centering
\includegraphics[width=\columnwidth ]{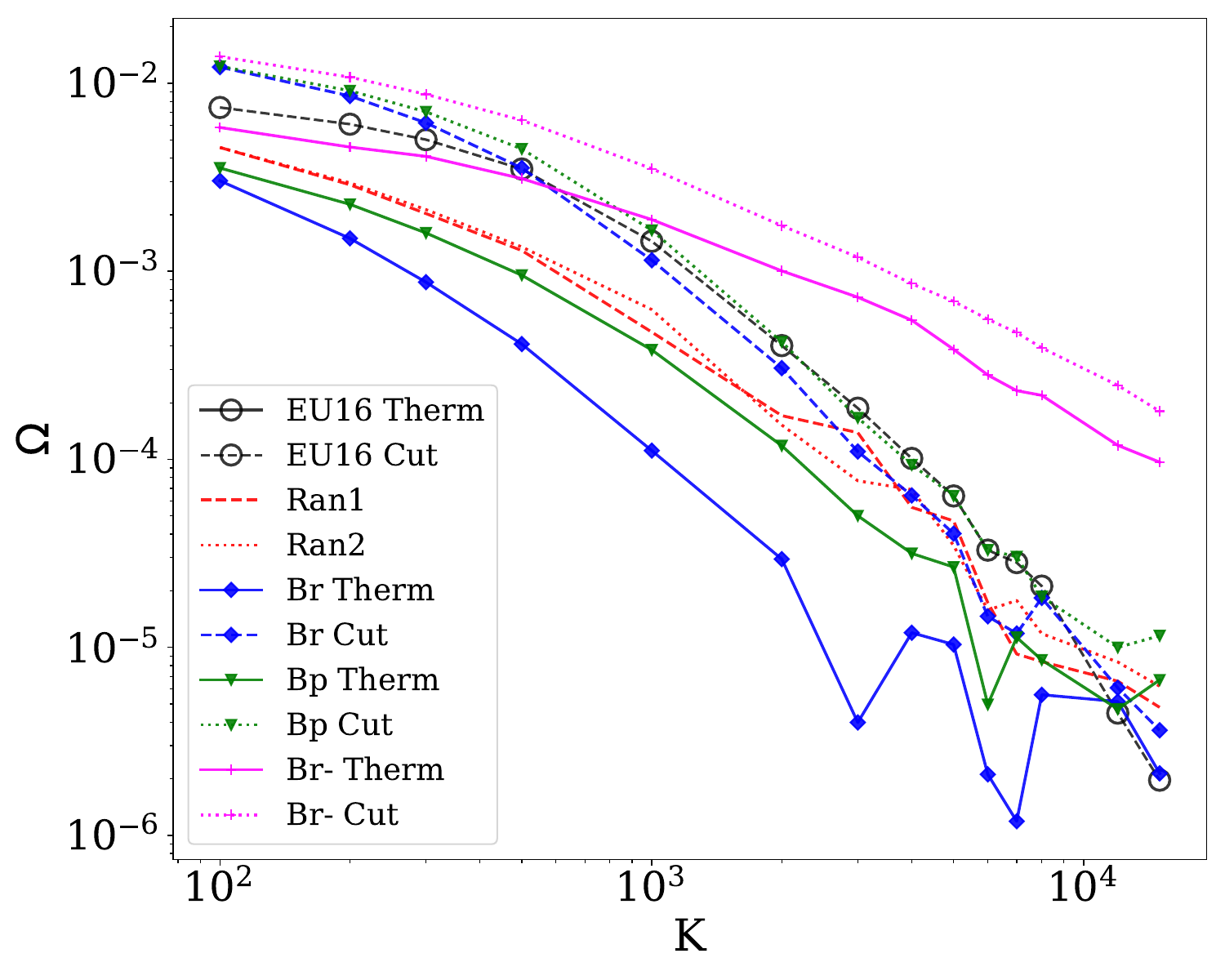}
    \caption{Comparison of numerical solutions of $\Omega$ at the end of the thermalization (Therm), and after the cascade failures (Cut), in the steady states, for the original (EU16), randomly extended versions (Ran1, Ran2), bridged (Br), bypassed (Bp) and bridge truncated (Br-) EU16 networks, using $\alpha=3.0$. Bridge removals increase the global frequency spread, all the other solutions shrink it.}
    \label{fig:O-K3}
\end{figure}

Fig.\ref{fig:O-K3} shows the frequency spread results, for $\alpha=3$, where an increase of $\Omega$ for every modified networks considered is depicted. The increase is relatively small, except for the bridge removal cases and negligible for bridge additions. We show here results separately after the thermalization (Therm), and after the cascade failure (Cut), on the EU16 network. In general, we can observe an increase of $\Omega$ after the cascades. All modifications, except bridge removal increase efficiency, the best results are obtained by bridge addition.

The next figure is related to the EU22 power-grid. Fig. \ref{fig:22-O-K-a3} shows the frequency spread, using $\alpha=3.0$. Here the bypasses and the random link addition provide the best outcomes, while bridge addition does not results in an improvement. The bridge removal causes an increase in the $\Omega$ order parameter. The inset shows the deviations of varied network results with respect to the original EU22.

\begin{figure}
    \centering
\includegraphics[width=\columnwidth ]{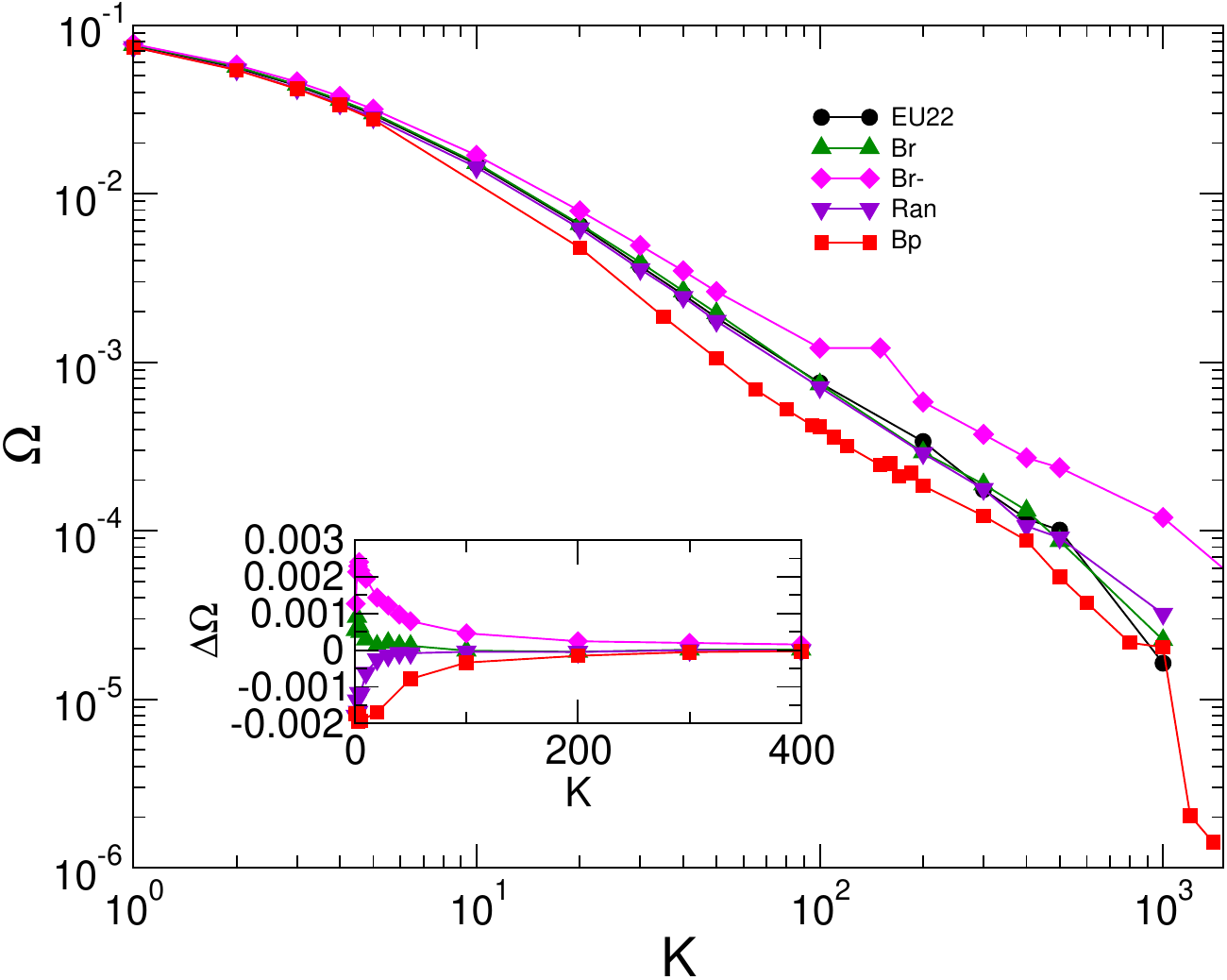}
    \caption{Comparison of numerical solutions of $\Omega$ at the end of the thermalization (Therm), and after the cascade failures(Cut), in the steady states, for the original (EU22), randomly extended (Ran1), bridged (Br), bypassed (Bp) and bridge truncated (Br-) EU22 networks, using $\alpha=3.0$. Here, bridge changes increase the global frequency spread, all the others shrink it, as one can see in the inset.}
    \label{fig:22-O-K-a3}
\end{figure}

\end{document}